\pdfoutput=0
\documentclass[11pt,preprint]{aastex}

\usepackage{natbib}
\usepackage{graphicx}
\usepackage{amsmath,amssymb}
\usepackage{url}
\usepackage{fancyref}
\usepackage[para]{footmisc}

\newcommand{\amm}{NH$_3$}

\shorttitle{Galactic Center}
\shortauthors{Elisabeth Mills}

\begin{document}

\title{VLASSICK: The VLA Sky Survey in the Central Kiloparsec}

\author{E.A.C. Mills\altaffilmark{1}, A.Ginsburg\altaffilmark{2}, J.M.D. Kruijssen\altaffilmark{3}, L. Sjouwerman\altaffilmark{1}, C.C. Lang\altaffilmark{4}, S.A. Mao\altaffilmark{1,5}, A. Walsh\altaffilmark{6}, M. Su\altaffilmark{7}, S.N. Longmore\altaffilmark{8}, J-H. Zhao\altaffilmark{9}, D. Meier\altaffilmark{10}, M.R.Morris\altaffilmark{11}}
\altaffiltext{1}{NRAO}
\altaffiltext{2}{ESO}
\altaffiltext{3}{MPA}
\altaffiltext{4}{U. Iowa}
\altaffiltext{5}{U. Wisconsin}
\altaffiltext{6}{Curtin}
\altaffiltext{7}{MIT}
\altaffiltext{8}{Liverpool John Moores}
\altaffiltext{9}{Harvard Smithsonian CfA}
\altaffiltext{10}{New Mexico Tech}
\altaffiltext{11}{UCLA}

\begin{abstract}
\vspace{-0.6cm}
At a distance of 8 kpc, the center of our Galaxy is the nearest galactic nucleus, and has been the subject of numerous key projects undertaken by great observatories such as Chandra, Spitzer, and Herschel. However, there are still no surveys of molecular gas properties in the Galactic center with less than 30\arcsec\, (1 pc) resolution. There is also no sensitive polarization survey of this region, despite numerous nonthermal magnetic features apparently unique to the central 300 parsecs. In this paper, we outline the potential the VLASS has to fill this gap. We assess multiple considerations in observing the Galactic center, and recommend {\bf a C-band survey with 10 $\mu$Jy continuum RMS and sensitive to molecular gas with densities greater than 10$^4$ cm$^{-3}$, covering 17 square degrees in both DnC and CnB configurations ($\theta_{res}\sim5\arcsec$), totaling 750 hours of observing time}. Ultimately, we wish to note that the upgraded VLA is not just optimized for fast continuum surveys, but has a powerful correlator capable of simultaneously observing continuum emission and dozens of molecular and recombination lines. This is an enormous strength that should be fully exploited and highlighted by the VLASS, and which is ideally suited for surveying the center of our Galaxy. 

\end{abstract}

\section{Introduction} 
\vspace{-0.3cm}
At a distance of $\sim8$ kpc, our Galactic center is by far the nearest galactic nucleus, making it possible to probe the components and processes that dominate here at scales (5\arcsec = 0.2 pc) that are impossible to resolve in other Galactic centers.  Much of what we now know about the Galaxy's Central kiloparsec  (hereafter, the CK) comes from surveys at a range of wavelengths using facilities such as Hubble, Chandra, Spitzer, Herschel, Fermi, and numerous ground-based facilities. We know that it is a reservoir of hot, dense and turbulent molecular gas, that the present-day star formation in this gas is apparently low \citep[e.g.][]{Longmore13}, but in the recent past it was high enough to form a large population of massive young stars and clusters \citep[e.g.,][]{Figer04,Mauerhan10}, and that either this star formation or the supermassive black hole may support enormous gamma-ray-emitting outflow lobes \citep[the `Fermi Bubbles', e.g.,][]{SSF10,Crocker12}. However, the current picture of this region is still very incomplete with regard to the dynamics, the 3D distribution of the dense gas, and the magnetic field.  

There have been no uniform surveys of the molecular gas in the Galactic center with better than 30\arcsec\, (1 pc) resolution, meaning that our knowledge of the gas conditions is limited to the average properties of giant molecular clouds, as opposed to the properties of clumps and cores, the direct precursors to forming individual stars. The majority of the studies of molecular gas in this region, e.g., HCN \citep{Jackson96}, 3 and 7 mm line surveys \citep{Jones12,Jones13} and \amm\, (Ott et al. submitted), are also confined to the central $<300$ parsecs, limiting our understanding of how the detailed kinematics and gas conditions in the Galactic center transition from those in the Galactic disk. Additionally, there are insufficient sensitive polarization studies in the CK  for a robust census of the magnetic field properties in this region. Thus at present, we lack the data necessary to answer key questions that remain about the center of our Galaxy, and galaxies in general: Does star formation proceed differently in extreme environments of galactic nuclei, compared to galactic disks? How is gas fed into the nucleus, and how is it ejected? How do magnetic fields influence all of these processes?

With the upgrade of the VLA, there is now a unique opportunity to fill this knowledge gap.
The new VLA correlator offers unprecedented sensitivity and wide instantaneous bandwidths that allow for fast mapping of large areas in both full stokes polarization imaging of continuum emission (tracing nonthermal features and the magnetic field properties) and spectral line imaging of numerous molecular and recombination lines (tracing the distribution and kinematics of the gas) at resolutions of a few arcseconds/ tenths of a parsec.  

However, PI-driven science cannot take full advantage of these opportunities. The time available in each of the hybrid configurations necessary to observe this low-declination region is limited to 18 days every 1.3 years, and the LST of the Galactic center is generally heavily oversubscribed (e.g., a factor of 5 in the last CnB configuration). As a result, science is being done piecemeal. The wealth of correlator configurations now possible leads to ever more inhomogeneous data sets that are not optimal for comparing the center of our Galaxy to the central kiloparsec of other galaxies, and as a result questions requiring a complete census of the CK remain unanswerable. 

Making a deep observation of the central kiloparsec a key project in the VLASS would yield a unique, homogenous data set ideal for studying the detailed physical processes in the nuclear bar of our Galaxy and for applying this understanding to extragalactic nuclei. The resulting survey data would not only address whether the extreme environment of the Galactic center affects star formation in this region, but would follow the full dynamic lifecycle of gas in a Galactic nucleus, from infall to outflow, and would help reveal the role of magnetic fields in all of these processes. The legacy value of this survey would be enormous: the largest and deepest radio continuum polarization image of the Milky Way nucleus, as well as the first high-resolution data set of a dense gas tracer in the CK, would serve as an invaluable pathfinder to define future higher-resolution studies of gas and continuum in the CK with the VLA, ALMA, and SKA. Below, we highlight a subset of the science that would be possible with a comprehensive survey of this region. 

\begin{figure}[p]
\hspace{0.2cm}
\includegraphics[scale=0.5]{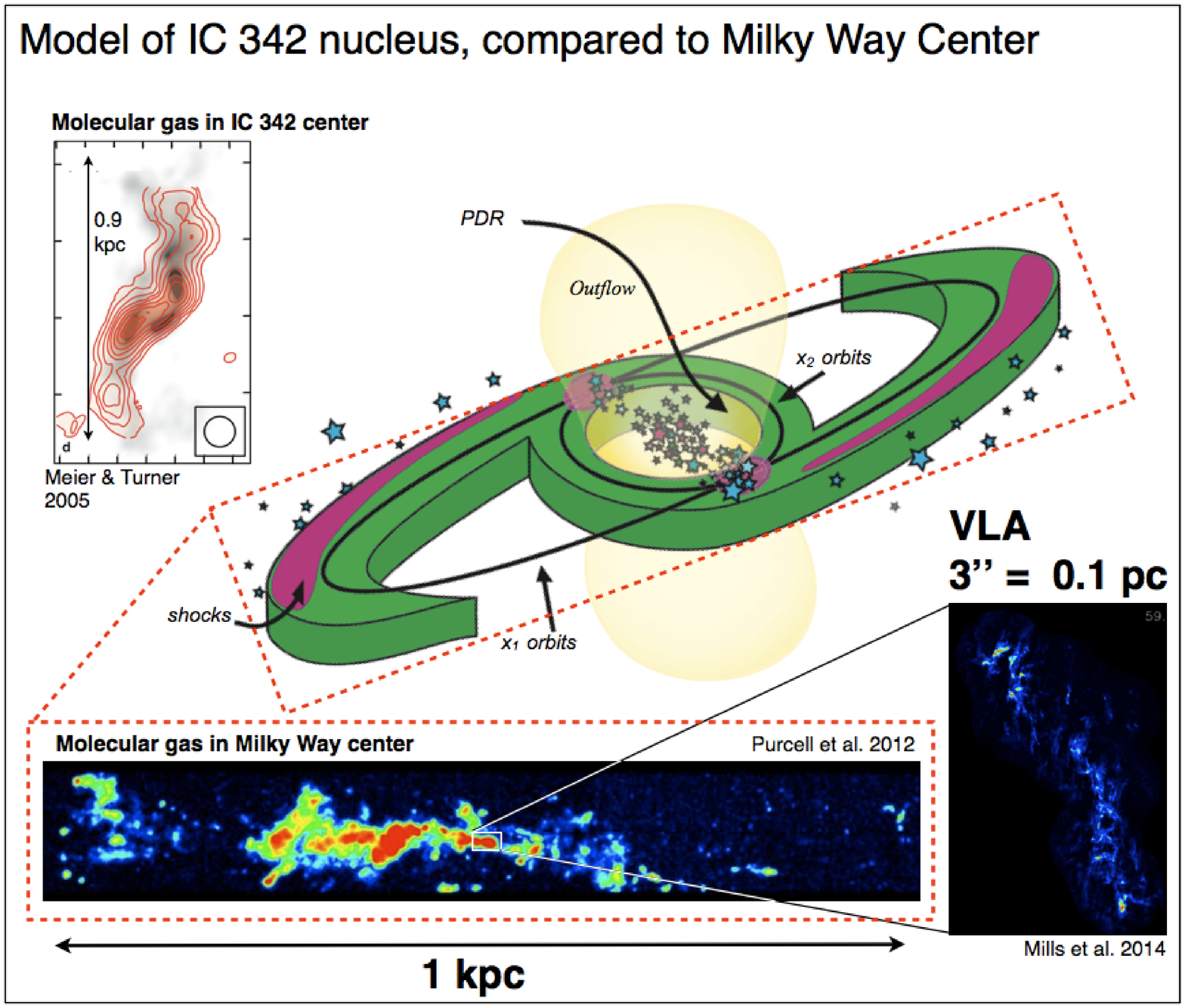}
\vspace{-0.5cm}
\caption{\em{A schematic (adapted from \citealt{Meier05}) showing the structure of a typical nearby nucleus (IC 342), based on observations of resolved extragalactic molecular line emission. The central kiloparsec of our Galaxy, shown here in \amm\, from \cite{Purcell12}, is suggested to have a similar (barred) structure, in which the gas follows x1 orbits along the bar until these orbits become self-intersecting near the inner Lindblad resonance \citep[R$\sim$0.5 kpc,][]{Binney91}, and the gas is shocked, becoming primarily molecular, and is funneled inward to a second class of stable orbits perpendicular to the bar at smaller radii (x2 orbits). Most surveys of the Galactic center have concentrated on the x2 orbits, and as a result much less is known about the evolution in gas properties as it transitions from the disk to the nucleus (e.g., near the x1-x2 orbit transition). 
}}
\end{figure}

\vspace{-0.5cm}
\section{Science Questions}
\vspace{-0.3cm}
\subsection{An Extreme Star Formation Environment}	
\vspace{-0.3cm}
The center of our Galaxy is an excellent analog to high-redshift star-forming regions \citep{KL13}. However, although bulk CK gas properties (e.g., kinematics, temperatures, and densities) are similar to those observed \citep{Genzel11} and simulated \citep{Dekel09} at high redshifts, the CK is deficient in star formation, violating `laws' that appear to hold in our Galaxy and others \citep{Kennicutt98, Lada12, Longmore13, Kruijssen13}. Explaining this discrepancy is then critical both for understanding our nucleus and for determining the proper star formation `laws' that apply to extreme extragalactic environments. 

{\bf Conditions for Star Formation: }
One goal would be to measure gas properties in the CK on the scales most relevant to forming individual stars: clumps and cores. Turbulence will be measurable from linewidths with any sufficiently abundant gas tracer, while temperatures and densities require a more specialized survey of specific tracers (e.g., NH$_3$ in K-band, H$_2$CO in C and Ku bands). In addition to measuring the turbulence of the molecular gas, modeling depolarization and structure functions of Faraday depths can be used to extract the turbulence properties of the plasma. Full polarization observations also allow Zeeman measurements of maser lines to infer the magnetic field strength in the molecular gas \citep[e.g.,][]{Crutcher10,Roberts99}. 

{\bf Quantifying Galactic center star formation: }
A variety of masers (e.g., CH$_3$OH at 6 and 12 GHz; H$_2$O at 22 GHz) can be used to address whether the apparent lack of maser emission (one potential tracer of massive star formation) in the CK is a result of resolution and sensitivity limitations of previous surveys \citep{Longmore13}. A sensitive continuum survey will also yield a complete sample of the upper-end of the initial mass function for young massive stars in the CK (via detection of ultra- and hypercompact HII regions and stellar wind sources), enabling the determination of a more robust star-formation rate directly from stellar number counts. 

\vspace{-0.5cm}
\subsection{Infall: Mapping the Detailed Kinematics of Gas from the Bar to the Black Hole} 
\vspace{-0.3cm}
Unlike many other nearby galaxies (e.g., IC342 in Figure 1), we must view the center of our own Galaxy edge-on. This greatly complicates our understanding of the true 3D structure of this region. In order to effectively compare gas properties in our Galactic center to those observed in others, we need to know where the gas is and how it is moving. One of the most interesting questions to answer from such a comparison is how gas flows from a galaxy's disk into the nucleus, ultimately feeding the black hole and periodic nuclear starbursts. 

By surveying both molecular and ionized gas over the entire CK it will be possible to refine and expand 3D models of the CK, e.g., resolving discrepancies in the model of the 100 parsec ring (Kruijssen, Dale, \& Longmore, in prep), which is suggested to be gas on the innermost x2 orbits \citep{Molinari11}. With a high resolution survey of this entire region, one can further incorporate variations in detailed physical properties into this model, for example, the clumping properties or turbulent density spectrum of the gas. 

\vspace{-0.5cm}
\subsection{Galactic Center Outflows from Small to Large Scales}
\vspace{-0.1cm}
The CK appears to be the launching point of outflows, e.g., a wind traced by high-velocity HI clouds \citep{McCG13}, and the `Fermi Lobes' \citep{SSF10}, analogues of which cannot at present be detected in other galaxies, but which may be common if \citep[as suggested by][]{Crocker12} they are driven by relatively normal nuclear star formation. One way to better understand the nature of these outflows and the mechanisms that support them is to connect the properties of smaller scale structures: putative jets from Sgr A* \citep{Li13} and new filamentary structures radiating from Sgr A \citep{Morris13}, to the larger structures:  the Galactic center lobe-- \citep[][, Figure 3]{SH85}, and jets in the Fermi bubbles \citep{SF12}. 

A primary goal would be to measure and compare the polarization properties and possibly the kinematics (via stacking of multiple recombination lines where the emission is sufficiently strong) of putative outflow features in the central degrees of the Galaxy, connecting the large-scale features to the smallest scales on which jets are suggested to emanate from Sgr A*. Combined with the detailed morphologies of these features that this survey would yield, these measurements would enable these structures to be physically associated, and for more precise expansion rates to be determined, constraining models for powering outflows from the CK. 

\vspace{-0.5cm}
\subsection {Large-scale Magnetic Field Strength and Geometry in the Galactic Center}
Some of the most striking radio features of the CK region are synchrotron-emitting filaments, linear over tens of parsecs scales (Figure 2) that are understood to be magnetic flux tubes infused with relativistic electrons streaming along an ordered, strong, poloidal field \citep{YZMC84}. More recent studies of the region have uncovered a population of fainter, shorter, and more numerous filamentary structures that may trace a more complicated magnetic field structure \citep[e.g.,][]{LaRosa04,YZ04}

Currently, only individual pointed studies of the magnetic filaments have been carried out, leaving the polarized nature of many of these features yet to be confirmed. Full Stokes polarization imaging, ideally to sensitivities of a few $\mu$Jy, is necessary to detect the full range of magnetized filamentary features over the wide range of brightnesses they exhibit. A sensitive polarization survey over a large range in $l$ and $b$ will constrain the full extent and geometry of the magnetic field in our Galaxy's nucleus, and will allow intensive study of the properties of the magnetic filaments and their environments to estimate their synchrotron lifetimes, to identify the site(s) of relativistic electron injection, and to gain a better understanding of the electron acceleration mechanism. 

Finally, by extending any CK survey at least $1\degr$ out of the plane, it will be possible to search for links between known magnetic filaments and fainter extensions at higher Galactic latitudes that may connect to larger-scale polarized lobes attributed to outflows from the center. Measuring the polarization properties in this transition region would clarify the role the magnetic field plays in shaping these large-scale vertical structures and controlling the flow of energy from the CK region.

\begin{figure}[p]
\hspace{0.2cm}
\includegraphics[scale=0.5]{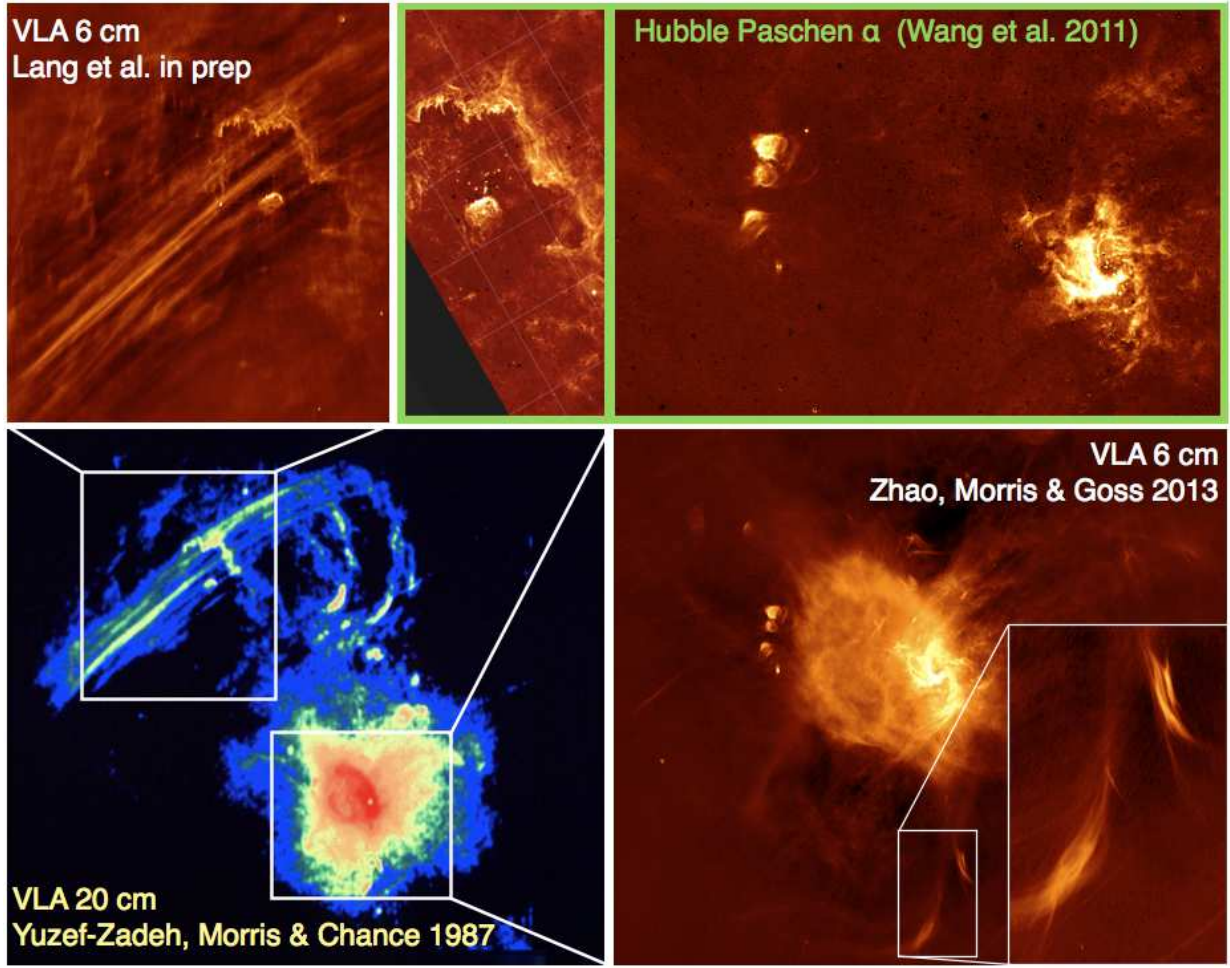}
\vspace{-0.5cm}
\caption{\em{{\bf Bottom Left:} 20 cm discovery image of radio features in the Galactic center, featured on the cover of Nature in 1984 \citep{YZMC84}.  {\bf Bottom right:} 6 cm image of Sgr A, using the new correlator capabilities of the VLA, and illustrating a wealth of complex new features: including new point sources, filamentary substructure and diffuse emission, that can now be detected in this region. {\bf Top Right:} For comparison, Hubble Paschen-$\alpha$ emission toward the same region, illustrating that the VLA data are not only comparable, but contain copious additional information. {\bf Top Left:} New 6 cm image of the Sickle HII region and nonthermal filaments, compared to a Paschen-$\alpha$ image of the same region. 
}}
\end{figure}

\vspace{-0.5cm}
\section{Technical considerations on Survey Design}
\vspace{-0.3cm}
\noindent{\bf Array Configuration:} At the declination of the Galactic center ($-30\degr$), hybrid array configurations are necessary for optimal image sensitivity and  fidelity (reconstructing the morphology of diffuse and extended features). While previous experience shows that imaging the complicated continuum substructures in this region is best done with a synthesis of at least 2-3 array configurations, molecular line science (apart from maser sources) depends primarily on brightness sensitivity, for which the most compact configurations (DnC and CnB) are most useful. 
		
\noindent{\bf Imaging:} Recent VLA surveys of the Galactic center (Multi-array imaging of Sgr A with an RMS of 10 $\mu$Jy, \citealt{Zhao13}; Multi-frequency polarization of the Radio arc, Lang et al. in prep; K-band studies in a sample of CK clouds, Mills et al. 2014) have already demonstrated successful image reconstruction in this complicated region, with dynamic ranges up to 100,000. Optimal image reconstruction of (non-masing) molecular line transitions will require combination with single-dish observations of the same area in order to recover emission on spatial scales greater than tens of arcseconds to a few arcminutes (depending on the chosen frequency). This has been successfully demonstrated for a Cycle 0 ALMA mosaic of one CK cloud (Rathborne et al. submitted) and will be further tested for VLA and GBT data in the central 10 parsecs (Mills \& Liu, in prep).

\noindent{\bf Polarization:} Detecting faint Galactic center nonthermal filaments \citep[typical spectral indices of $-0.5$,][]{LaRosa00} requires low frequencies to be optimally sensitive to synchrotron emission. In contrast, for rotation measure (RM) studies, higher frequencies yield better sensitivity to large values of $|$RM$|$ as well as extended structure in Faraday depth space. However, the RM uncertainty increases with frequency, as it is related to (1/($\lambda_{max}^2-\lambda_{min}^2$)). Another consideration is `off-axis' linear polarization in images containing the bright Sgr A complex just at the edge or outside of the primary beam. This false polarization is significant (for Sgr A*, upwards of 50\%) and can mask low-level linear and circular polarization in the center of the proposed survey region \citep{Bower02}. This effect is minimized at frequencies higher than S-band, where the smaller beam limits the region in which Sgr A is strongly detected in the sidelobes. Together, these considerations favor a survey at intermediate frequencies (e.g., C or X-band). 		
		
\noindent{\bf Sensitivity:} Efficient mapping of the CK gas requires an abundant molecule (e.g., H$_2$CO, OH, NH$_3$). The molecular line sensitivity will be driven by a desire to get detailed kinematics of the gas (velocity resolutions $<2$ km s$^{-1}$) and trace the cascade of turbulent energy to small scales and correspondingly small linewidths. In contrast, the continuum sensitivity is driven by measuring the polarization of faint nonthermal filaments, for which an RMS of a few $\mu$Jy is desired for a full census of their properties. Further, avoidance of RFI requires frequencies $>2.4$ GHz to achieve a large bandwidth for efficiently yielding this sensitivity. However, high frequencies have small primary beams that are impractical for mosaicing large regions.

\noindent{\bf Spatial Resolution:} Resolution on the order of a few arcseconds or tenths of a parsec is necessary for optimal sensitivity to compact sources of maser and continuum emission, especially the nonthermal filamentary features, many of which are still unresolved at arcsecond resolution. It is also crucial for resolving the internal structure of molecular clouds at much better than the tidal radius (a few pc) to compare with kinematic models of gas in the region and understand why most of this gas does not appear to be forming stars. While lower frequencies have the benefit of increased sensitivity to emission on large scales, which may reduce the need for complementary single-dish observations of the molecular gas, they limit the resolution achievable with a survey or follow-up. 

\noindent{\bf Survey Area:} Covering the CK (including gas on both x1 and x2 orbits) requires longitudinal coverage of $l=\pm3.5\degr$ for comparison to other galactic nuclei and tracing the kinematics of infalling of gas from the inner Lindblad resonance to the supermassive black hole, including the unusual x1 cloud Bania's Clump 2 \citep{Bania77,Stark86}. A latitudinal coverage of $b=\pm1.0\degr$ will cover almost the entirety of the molecular gas; increasing this to $b=\pm1.5\degr$ in the central regions would cover high-latitude extensions of the nonthermal filaments, and would trace structures that may be related to large scale outflows, e.g., the Galactic center lobe \citep{SH85} or putative X-ray counterparts of the Fermi Jets, extending $\sim1.5\degr$ from Sgr A* \citep{Nakashima13}. 

\noindent{\bf Choice of frequency:} 
\vspace{-0.5cm}
\begin{itemize}
	\item {\bf L band}, given the limited RFI-free bandwidth, is less sensitive than other bands, though this is partially compensated for by the large available field of view for mapping. It cannot achieve better than $1''$ resolution, and the large field of view also makes it more susceptible to off-axis polarization contamination. The four OH lines in this band are a good tracer of the low-density molecular gas. 
	\vspace{-0.2cm}
	\item {\bf S band} offers a larger field of view and more comparable sensitivity, and includes the $^2\Pi_{1/2}$ $J=12$ transitions of CH, potentially one of the best tracers of the low-density gas in this region \citep{Magnani06}.
	\vspace{-0.2cm}
	\item {\bf C band} is a reasonable compromise for sensitivity to nonthermal and thermal sources of emission. It includes both the ubiquitous 6.6 GHz CH$_3$OH radiatively-excited maser and the $1_{10}-1_{11}$ transition of H$_2$CO, a good tracer of moderate to high density gas, which would be observed primarily in absorption against the CMB and other continuum emission.
	\vspace{-0.2cm}
	\item {\bf X band} is still sensitive to nonthermal emission, however with the same total available bandwidth and a smaller beam than C-band, it is less efficient for mapping large areas. The 1--0 line of HC$_3$N lies in this band: a good tracer of moderately dense gas, and under some conditions a weak maser \citep{Hunt99}.
	\vspace{-0.2cm}
	\item {\bf Ku band} is the highest frequency at which one does not incur significant extra overhead from pointing. It is even less efficient for mapping large areas than X band. There are no fundamental rotational transitions of common molecules in this band, though there are the $2_{11}-2_{12}$ transition of H$_2$CO and the radiatively excited CH$_3$OH maser line at 12 GHz. 
	\vspace{-0.2cm}
	\item {\bf K band} is optimal for coverage of (optically-thick) thermal free-free emission, and covers multiple strong ammonia transitions that can be used to derive gas temperatures. However, it is significantly less sensitive to synchrotron sources, and the reduced field of view and increased overhead time makes mapping more than the innermost $\sim1\degr$ a several thousand hour endeavor that is dependent upon good weather conditions.
\end{itemize}

\vspace{-0.5cm}
\section{Recommendation}
\vspace{-0.1cm}
{\bf We have identified the C band (4-8 GHz) as the optimal frequency for a large-area survey of the nuclear bar and inner bulge of our Galaxy. } As detailed above, C band is optimal for efficient coverage of a large survey area, inclusion of strong molecular line features (H$_2$CO, a ubiquitous, though not optically-thick, absorption line tracing moderate to high density gas), and sensitivity to nonthermal emission.  A strength of this setup is that it should be synergistic with other potential components of a VLA sky survey. We note that many of the driving goals for this choice of frequency (strong molecular lines, optimal areal coverage and continuum sensitivity) will also be ideal for surveys of the Galactic plane and nearby galaxies. We desire to cover a region of 17 square degrees (Figure 3), which will include the entire central kiloparsec with a latitudinal coverage of $b$ = $\pm1.0\degr$, and increasing to $b$ = $\pm1.5\degr$ in the central region in order to include candidate outflow structures.

Survey sensitivity is driven by the goal of detecting H$_2$CO in all gas with n $>2\times10^4$ cm$^{-3}$ (expected brightness temperature of -1.6 K) with a S/N $>$5 per 5  km s$^{-1}$ channel in DnC configuration. As CnB-configuration has much lower surface brightness sensitivity, here our goal is only to detect absorption where there is a background illumination source brighter than about 10 mJy. This includes about 75\% of the inner 300 pc \citep{Law08}. Obtaining these line sensitivities will also yield $\sim10 \mu$Jy continuum sensitivity, sufficient to measure polarization in newly-discovered weak filamentary structures like those recently detected in Sgr A \citep{Morris13}. 




Observing in both DnC and CnB configurations will yield resolutions of 5\arcsec\, ($\sim$0.2 pc) a size scale comparable to those of ultra/hypercompact HII regions and approaching the resolution of individual cores in the molecular clouds, while retaining sensitivity to sources on scales as large as 300$''$ (10 pc). Ideally, at least the formaldehyde would be supplemented with single-dish (GBT) observations, as much of the gas emission arises on these larger scales. If desired, zero-spacing information for the continuum measurements could be provided by existing GBT data \citep{Law08} or upcoming CBASS data \citep{King13}.  

To accomplish these goals with the VLA  would require {\bf 750 hours} total, including estimated overheads of 25\% for VLASS surveying modes. The data rate required to observe the continuum, 20 recombination lines,  H$_2$CO and the CH$_3$OH maser is 15.7 MB/s, which will then ultimately result in a raw data volume for this survey of 34 TB.


\vspace{-0.5cm}
\section{Possible Alternatives}
\vspace{-0.3cm}
The simplest way to reduce the time required for this survey is to reduce the area. For example, a 10 square degree area with a latitudinal coverage of $b$ = $\pm0.5\degr$ ($b$ = $\pm1.0\degr$ in the central three degrees) would have the same longitudinal coverage, and would take {\bf $\sim$450 hours}. It could achieve the majority of the science goals as it would still cover the bulk of the molecular gas in the region, and would include several outflow features and high-latitude extensions of the nonthermal filaments. However, it would limit the ability to connect these features to the Fermi lobes (and jet). 

Further reducing the survey area, e.g., covering only the inner 2 degrees (overlapping with many smaller GC studies shown in Figure 3) would come at a cost of limiting the scope of the science: losing the opportunity to connect the magnetic field and gas properties in the CK to those in the disk and bulge, and to study outflows from this region. Limiting the survey to a single configuration (CnB or DnC) would either drastically reduce the sensitivity to extended gas structure, or decrease the survey's sensitivity to masers and compact continuum structure, yielding less complete statistics on the star formation in this region. Finally, reducing the sensitivity will progressively limit any ability to map gas kinematics, or to detect thermal gas emission at all, as well as decreasing the number of continuum features for which polarization can be measured, yielding a less complete picture of the magnetic field structure.

\noindent{\bf In summary}, a 4-8 GHz survey of the inner 3.5 degrees of our Galaxy in the DnC and CnB configurations at a sensitivity of 10 $\mu$Jy will provide:
\vspace{-0.3cm}
\begin{enumerate}
\item {\bf A complete census of recent ($<5$ Myr) star formation activity} allowing us to address whether the apparent lack of star formation in the CK is a result of resolution and sensitivity limitations of previous surveys.

\item {\bf Detailed kinematic structure of the dense gas} at a resolution well below the tidal radius which we can use to refine 3D and kinematic models of gas in the region and understand why most of this gas does not appear to be forming stars.

\item {\bf A physical connection between small and large outflow features} from the Sgr A* jet to the `Fermi Lobes', constraining models for powering outflows from the CK.

\item {\bf A map of the extent and geometry of the magnetic field in our Galaxy's nucleus}, clarifying the role the magnetic field plays in regulating star formation and controlling the flow of energy from the CK region.

\item {\bf Sensitive images of continuum, recombination and molecular lines} which will serve as the basis for numerous additional experiments, including:
\begin{itemize}
\vspace{-0.1cm}
\item time domain studies, e.g., ultracompact HII region flickering \citep{DePree13}, maser variability \citep{Goedhart04}, and the proper motion of continuum features \citep{Zhao13}
\vspace{-0.1cm}
\item constraints on the gas density via Ku-band followup of a subset of the gas in the  $2_{11}-2_{12}$ transition of H$_2$CO \citep{Ginsburg11}.
\vspace{-0.1cm}
\item studies of stellar remnants including X-ray binaries (expected F$_{\nu}$ $\sim$10 $\mu$Jy - a few mJy),  steep-spectrum point sources which may be candidate pulsars, and planetary nebulae, few of which have been previously detected in the survey area \citep[e.g.,][]{Parker06}.
\end{itemize}
\end{enumerate}

Ultimately, the VLASS has the potential to provide a unique data set that will finally make it possible to follow the mass flow and energy cycles that feed star formation and the supermassive black hole at the center of the Galaxy.



\begin{figure}[tbh]
\hspace{0.2cm}
\includegraphics[scale=0.32]{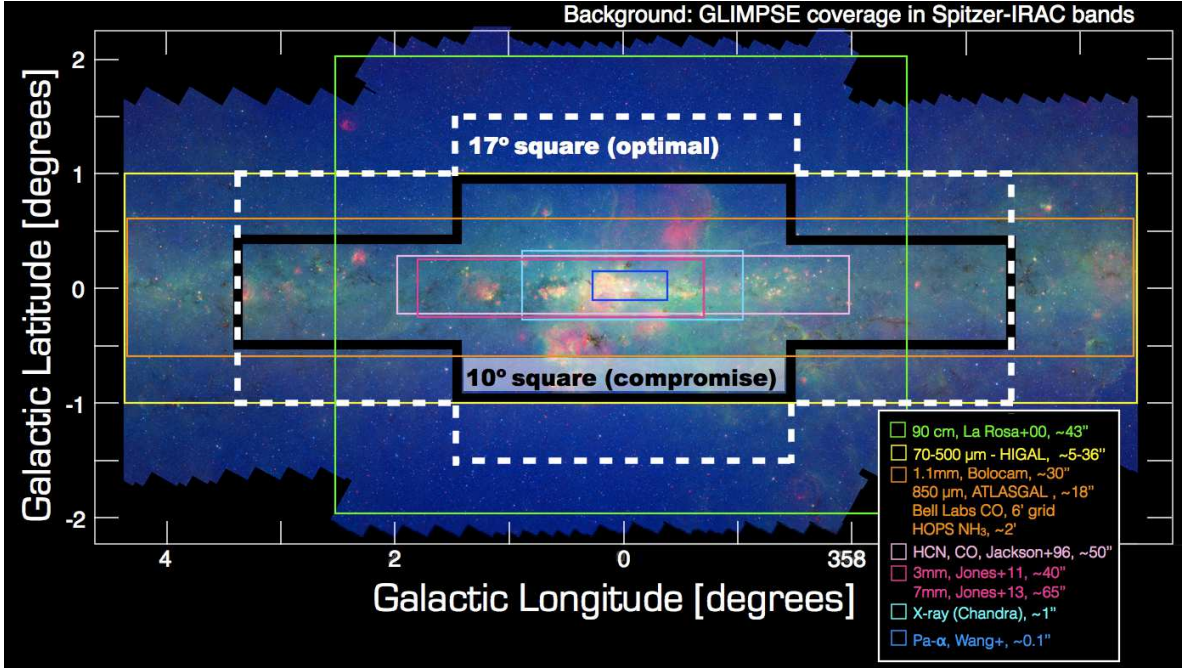}
\vspace{-0.3cm}
\caption{\em{ Recommended coverage areas for a deep VLASS mosaic of the Galactic center in C band (4-8 GHz) continuum, recombination lines, H$_2$CO, and CH$_3$OH masers, compared to coverage of existing surveys. The diffuse red emission at $l=359.5,b=0.5$ enclosed in a shell of green emission is the GC lobe \citep{SH85}, suggested to be an outflow from the Galaxy's nucleus.
}}
\end{figure}

\vspace{-0.5cm}
\section{Acknowledgements}
The authors would like to thank Daniel Wang, Namir Kassim, Juergen Ott, John Bally, Jill Rathborne, Farhad Yusef-Zadeh, Naomi McClure-Griffiths, and Katharine Johnston for useful discussion at the IAU 303 Galactic center symposium and for additional suggestions for the development of this white paper.  

\bibliographystyle{hapj}
\bibliography{whitepaper.bib}

\end{document}